\documentclass[cits]{PoS}
\pdfoutput=0
\title{More effects of Dirac low-mode removal}

\ShortTitle{More effects of Dirac low-mode removal}

\author{\speaker{M. Schr\"ock}\\
       Institut f\"ur Physik, FB Theoretische Physik, Universit\"at Graz\\
       E-mail: \email{mario.schroeck@uni-graz.at}}

\author{M. Denissenya\\
        Institut f\"ur Physik, FB Theoretische Physik, Universit\"at Graz\\
        E-mail: \email{mikhail.denissenya@uni-graz.at}}

\author{L.~Ya.~Glozman\\
        Institut f\"ur Physik, FB Theoretische Physik, Universit\"at Graz\\
        E-mail: \email{leonid.glozman@uni-graz.at}}
        
\author{C.~B.~Lang\\
        Institut f\"ur Physik, FB Theoretische Physik, Universit\"at Graz\\
        E-mail: \email{christian.lang@uni-graz.at}}        

\abstract{In previous studies we have shown that hadrons, except for a 
pion, survive the removal of the lowest lying Dirac eigenmodes from the 
valence quark propagators. The low-modes are tied to the dynamical breaking of chiral 
symmetry and we found chiral symmetry to be restored by means of 
matching masses of chiral partners, like, e.g., the vector and axial vector 
currents. Here we investigate the influence of removing the lowest part 
of the Dirac spectrum on the locality of the Dirac operator.
Moreover, we analyze the influence of low-mode truncation on the quark 
momenta and thereupon on the hadron spectrum and, finally,
introduce a reweighting scheme to extend the truncation to the sea quark sector.
}

\FullConference{31st International Symposium on Lattice Field Theory LATTICE 2013\\
		 July 29 -- August 3, 2013\\
		 Mainz, Germany}

\usepackage[tight]{units}
\usepackage{amsmath}

\newcommand{\eq}[1]{(\ref{#1})}
\newcommand{\Eq}[1]{Eq.~(\ref{#1})}
\newcommand{\fig}[1]{Fig.~{\ref{#1}}}

\newcommand{\Sec}[1]{Sec.~{\ref{#1}}}

\newcommand{\RD}[1]{{\mathrm{red}(#1)}}

\newcommand{\LM}[1]{{\mathrm{lm}(#1)}}

\newcommand{\be}{\begin{equation}}
\newcommand{\ee}{\end{equation}}

\newcommand{\bigO}{\mathcal{O}}
\newcommand{\e}{\text{\,e}}

\newcommand{\D}[1]{\mathcal{D}#1\,}
\newcommand{\lnorm}{\left\|}
\newcommand{\rnorm}{\right\|}

\begin{document}

\section{Motivation}

\begin{figure}[htb]
	\center
	\includegraphics[width=0.75\textwidth]{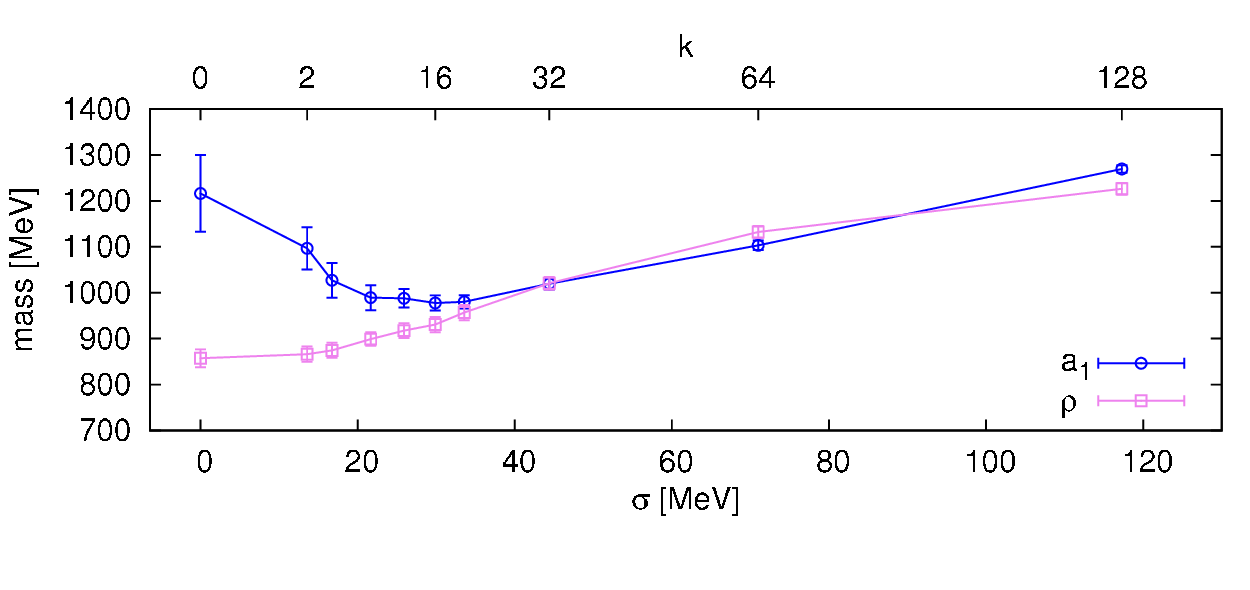}
	\caption{The masses of the vector and 
axial vector mesons, from left to right removing increasingly more and more Dirac low-modes.}
	\label{fig:fitted_masses_mesons}
\end{figure}

In previous work \cite{Lang:2011ai, Lang:2011vw, Schrock:2011hq, Glozman:2012fj}
we studied the effects of artificially removing the chiral condensate from the
valence quark sector.
We excluded a variable number of the lowest eigenmodes of the hermitian Dirac operator
$D_5\equiv \gamma_5 D$ from the valence quark propagators and subsequently investigated
the hadron spectrum within this approximation.
While the exponential decay in the pion channel got lost, the correlators of all 
other mesons and baryons we studied retained an exponentially decaying signal with 
significantly improved signal-to-noise ratios.
Therefore we were able to extract masses of the hadron states within our approximation.
In \fig{fig:fitted_masses_mesons} we show the masses of the vector and 
axial vector mesons as a function of the truncation level.

We observe degenerate masses of the would-be chiral partners $\rho$ and $a_1$ after having
subtracted modes up to $\sim\unit[30]{MeV}$ (twice the bare quark mass in our setup) 
of the spectrum 
of $D_5$ which indicates
the restoration of the dynamically broken chiral symmetry.
While the matching of the two masses in \fig{fig:fitted_masses_mesons} is expected, the increasing
behaviour is rather counter intuitive and will be explored here.

\begin{figure}[htb]
	\center
	\includegraphics[width=0.6\textwidth]{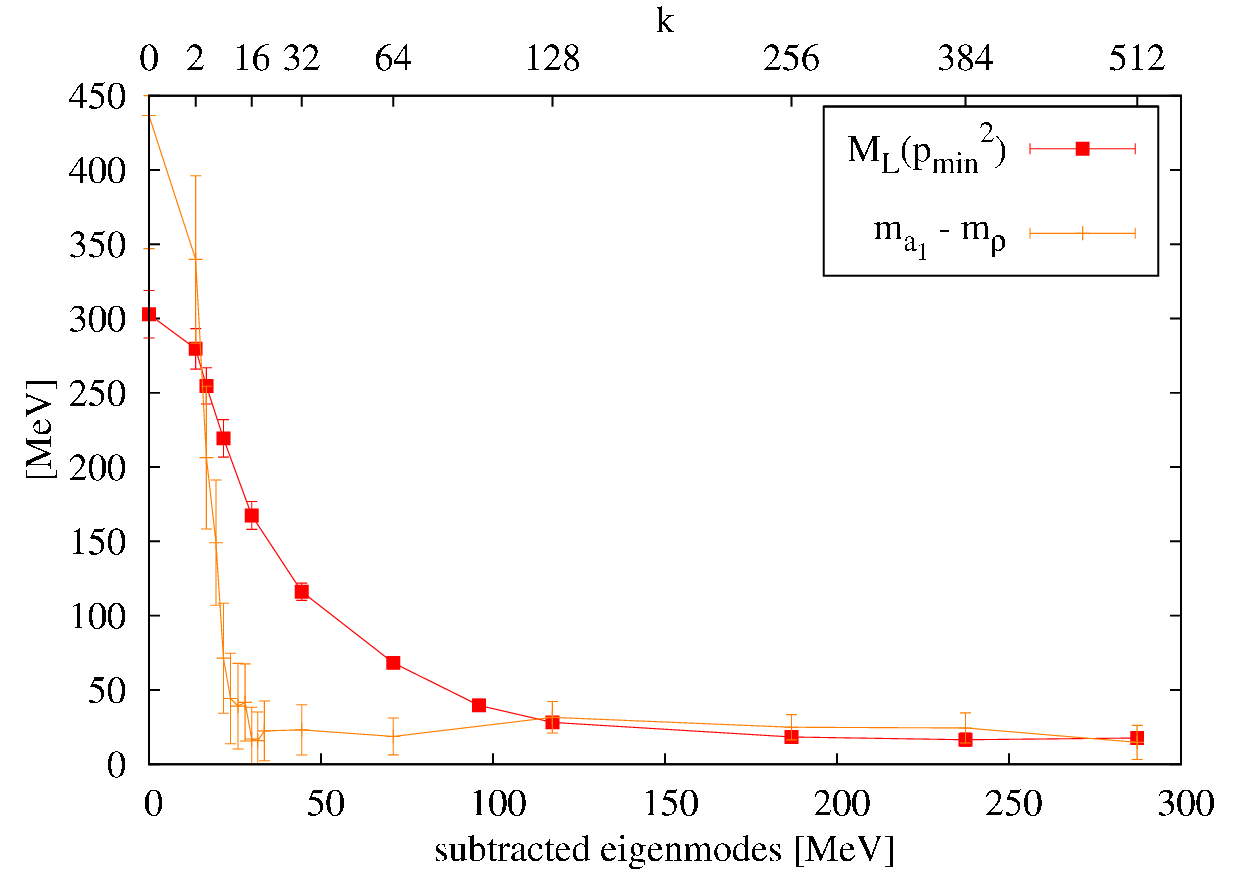}
	\caption{A comparison of the amount of dynamically generated mass in the quark mass
	function and the splitting of $\rho$ and $a_1$ as a function of the truncation level.}
	\label{fig:Mpmin}
\end{figure}

Another indication of a restored chiral symmetry is the loss of dynamically generated mass
in the quark propagator. In \cite{Schrock:2011hq} it has been shown that when removing the
low-lying part of the Dirac spectrum, the quark mass function (in Landau gauge) indeed
becomes flat.
When comparing the amount of dynamically generated mass as a function of the truncation level
with the splitting of the would-be chiral partners $\rho$ and $a_1$, see \fig{fig:Mpmin},
we obtain different truncation levels for the restoration of the chiral symmetry.
This discrepancy will be addressed here as well.

The remainder is organised as follows: in \Sec{sec:locality} we investigate the
violation of locality within our approximation and in \Sec{sec:massgen} the issue of increasing
hadron masses under Dirac low-mode truncation is studied. In \Sec{sec:seaquarks}
the possible extension of the approximation to the sea quark sector is explored 
before concluding in \Sec{sec:conclusions}.

\section{Locality properties of the truncated CI Dirac operator}\label{sec:locality}

Locality of the Dirac operator is a vital property for a quantum field theory 
since it ensures the causality of the theory. 
The original CI Dirac operator is, like the Wilson operator, \emph{ultra local} by definition.
The Neuberger overlap operator violates locality at finite 
lattice spacing \cite{Horvath:1998cm}.
The authors of \cite{Hernandez:1998et} have shown, though, that the nonlocal
contributions of the overlap operator fall  exponentially with the distance $r/a$
to the source point
and thus locality will eventually be restored in the naive continuum limit.
It is not clear a priori to what extent the exclusion of the lowest lying part of 
the Dirac eigenspectrum violates the locality of a Dirac operator like the CI operator.
The latter will be analyzed here.

We project out a single column of the hermitian Dirac operator
\be\label{eq:psiloc}
  \psi(x)^{\left[x_0,\alpha_0,a_0\right]} = \sum_{y}{D_5}(x,y)\,\eta(y)^{\left[x_0,\alpha_0,a_0\right]}\,,
\ee
where we use matrix/vector notation in color and Dirac space and the multiindex ${\left[x_0,\alpha_0,a_0\right]}$
labels a point-source
\be
  \eta_a^\alpha(y)^{\left[x_0,\alpha_0,a_0\right]} = \delta(y-x_0)\,\delta_{\alpha\alpha_0}\,\delta_{aa_0}\;.
\ee
In practice we set $x_0=0$ without loss of generality.
\Eq{eq:psiloc} defines a set of 12 columns of the Dirac matrix, 
one for each combination of color $a_0$ and Dirac  $\alpha_0$ indices.

Next, we define a function that serves as an upper bound of the contributions from $\psi(x)$
as a function of the distance $r$ to the source while taking the periodic boundary conditions of
the lattice into account:
\be\label{eq:fr}
  f(r) = \max_{x,\alpha_0,a_0}  \left\{ \lnorm \psi(x) \rnorm \,\left|\; \left| x \right| = r\right. \right\}\,,
\ee
where $\lnorm\cdot\rnorm$ is the usual vector norm over the internal color and Dirac structures of $\psi(x)$.
Then we can analyze the expectation value $\left<f(r)\right>$, which serves as a measure 
for the violation of locality.

We study \eq{eq:fr} for the low-mode truncated Dirac operator $D_5$, i.e., 
we consider columns of the truncated operator
\be
  \psi(x)^{\left[x_0,\alpha_0,a_0\right]}_\RD{k} = \sum_{y}D_5(x,y)\eta(y)^{\left[x_0,\alpha_0,a_0\right]} 
  - \sum_{i=0}^k \mu_i  v_i(x) \sum_{y}v_i(y)^\dagger \eta(y)^{\left[x_0,\alpha_0,a_0\right]}
\ee
where the $\mu_i$ are the eigenvalues of $D_5$ and $v_i$ the corresponding eigenvectors.
\begin{figure}[h]
\centering
\includegraphics[width=0.75\columnwidth]{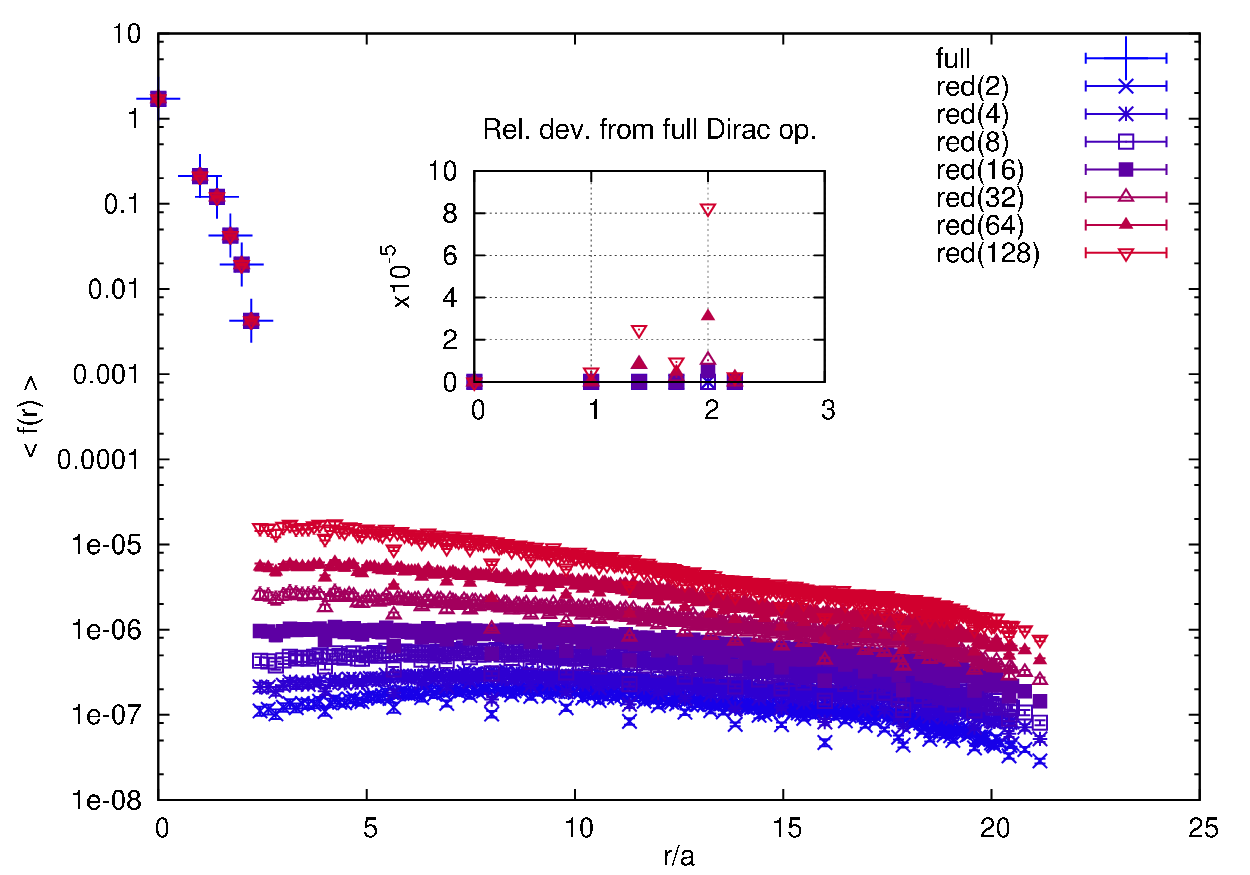}
\caption{The expectation value $\left<f(r)\right>$ for the truncated CI Dirac operator.
Small nonlocal contributions are present.
}\label{fig:red}
\end{figure}
From these we calculate $\left<f(r)\right>$ for truncation steps in powers of 2 from 
$k=2,\hdots,128$, see \fig{fig:red}.
First, we observe that the deviation from the nonzero contributions of the full
operator are of the order of $10^{-5}$, as can be seen from the inner plot of the figure, 
and thus are very small.
Moreover,  the truncated Dirac operator collects some nonlocal contributions of similar order, $10^{-8} - 10^{-5}$, from distant points on the lattice. 

This investigation shows that for our specific setup the
nonlocal contributions of the truncated Dirac operator are very small and the locality of the theory 
is conserved in good approximation.

\section{Mass generation}\label{sec:massgen}
In \cite{Schrock:2011hq} we investigated the quark propagator in Landau gauge\footnote{The code package \emph{cuLGT} \cite{Schrock:2012rm, Schrock:2012fj} for
gauge fixing on graphic processing units has been used to fix the gauge.}
under Dirac low-mode removal, with the goal of finding the effects on the quark
wavefunction renormalization function and on the quark mass function, which exhibits the
dynamical generation of mass due to the dynamical chiral symmetry breaking.
The first observation was that removing the lowest Dirac modes causes the dynamical mass
generation to cease, while the value of the bare mass (which is determined by the magnitude of the lowest
Dirac eigenmodes) is not affected.
The truncation level $k\approx 128$, at which the quark mass function appears flat, must coincide
with the level of complete removal of the chiral condensate from the valence quark sector.
Not only the mass function is affected by Dirac low-mode truncation,
the quark wavefunction renormalization function has also been found to be strongly 
suppressed when subtracting more and more low-modes.
Infrared suppression of the wavefunction renormalization function, which appears as 
an overall factor of the renormalized quark propagator, can be interpreted 
as suppression and eventual extinction
of low-momentum quarks in the spectrum.
The latter is in accordance with the fact that the low-momentum states of quarks 
are directly connected to low Dirac eigenvalues as can be derived explicitly 
for free quarks.

Matching of the masses of most of the chiral partners is found at an
earlier truncation level than suggested by the vanishing of the dynamically generated mass
of quarks: degenerate states of, e.g., the vector and axial vector currents are observed at
truncation level $k\approx 16$ while the chiral condensate decreases
with the truncation level until it disappears completely only at $k\approx 128$
as shown in \fig{fig:Mpmin}.
This can be explained by taking the increased momenta of the quarks under Dirac low-mode
removal into account. The early chiral restoration, as displayed by the degeneracy of states,
must be an effective restoration that is a combination of two underlying phenomena: 
first, the dynamically
generated mass of the quarks has shrunk to about sixty percent of its original value
and second, the momenta of the quarks are increased such that the effective dynamical mass at that
momentum value tends towards zero.

\section{The sea quark sector}\label{sec:seaquarks}

Throughout our work 
we used gauge configurations with two dynamical fermions and
only low-mode truncated the valence quark sector.
Here we elaborate how the sea quark sector can in principle be low-mode 
truncated \emph{a posteriori} via a reweighting procedure of the configurations.

We recall that we calculate observables $\bigO$ on the lattice via
\be\label{eq:pathI}
	\left<\bigO\left[U\right]\right> 
	= \frac{\int \D{U}   \,\e^{-S_G\left[U\right]} \det{D_u}\det{D_d} \bigO\left[U\right]}
		{ \int \D{U}  \,\e^{-S_G\left[U\right]}\det{D_u}\det{D_d}}\,.
\ee
In the latter expression the fermionic degrees of freedom have been integrated out
giving rise to the fermion determinant.
In the following we assume the two light quark flavors to be degenerate.

The fermion determinant can be written as the product of the Dirac eigenvalues $\lambda_i$
and therefore can be split into the product of a low-mode part and a reduced part
\be
	\det{D}_{\LM{k}} = \prod_{i\leq k}\lambda_i\,,\qquad
	\det{D}_{\RD{k}} = \prod_{i> k}\lambda_i\,.
\ee
Then we can formally include a weight $w_k$, which we define as
\be\label{eq:w}
	w_k\equiv \left(\det{D}_{\LM{k}}\right)^{-2},
\ee
into \eq{eq:pathI} in order to cancel the low-mode contribution of the fermion determinant
\begin{align}\label{rewI}\
	\left<\bigO\left[U\right]\right>_{w_k} 
	= \frac{\int \D{U}   \,\e^{-S_G\left[U\right]} \left(\det{D}_{\RD{k}}\right)^2 \bigO\left[U\right]}
		{ \int \D{U}  \,\e^{-S_G\left[U\right]}\left(\det{D}_{\RD{k}}\right)^2}.
\end{align}
Consequently, only the reduced part of the fermion determinant remains
in the path integral to represent the sea quarks.
Replacing the integration over the gauge fields in \eq{rewI} with a Monte Carlo 
integration, this is equivalent to
\be\label{OUwn}
	\left<\bigO\left[U\right]\right>_{w_k} \approx \frac{\sum_{n}\bigO\left[U_n\right]w_k\left[U_n\right]}{\sum_{n}w_k\left[U_n\right]}
\ee
where the finite number of gauge field configurations $U_n,\,n=1,\hdots,N,$ have been generated with the 
standard weight factor $\e^{-S_G\left[U\right]} \left(\det{D}\right)^2$.

\begin{figure}[h]
\centering
\includegraphics[width=0.49\columnwidth]{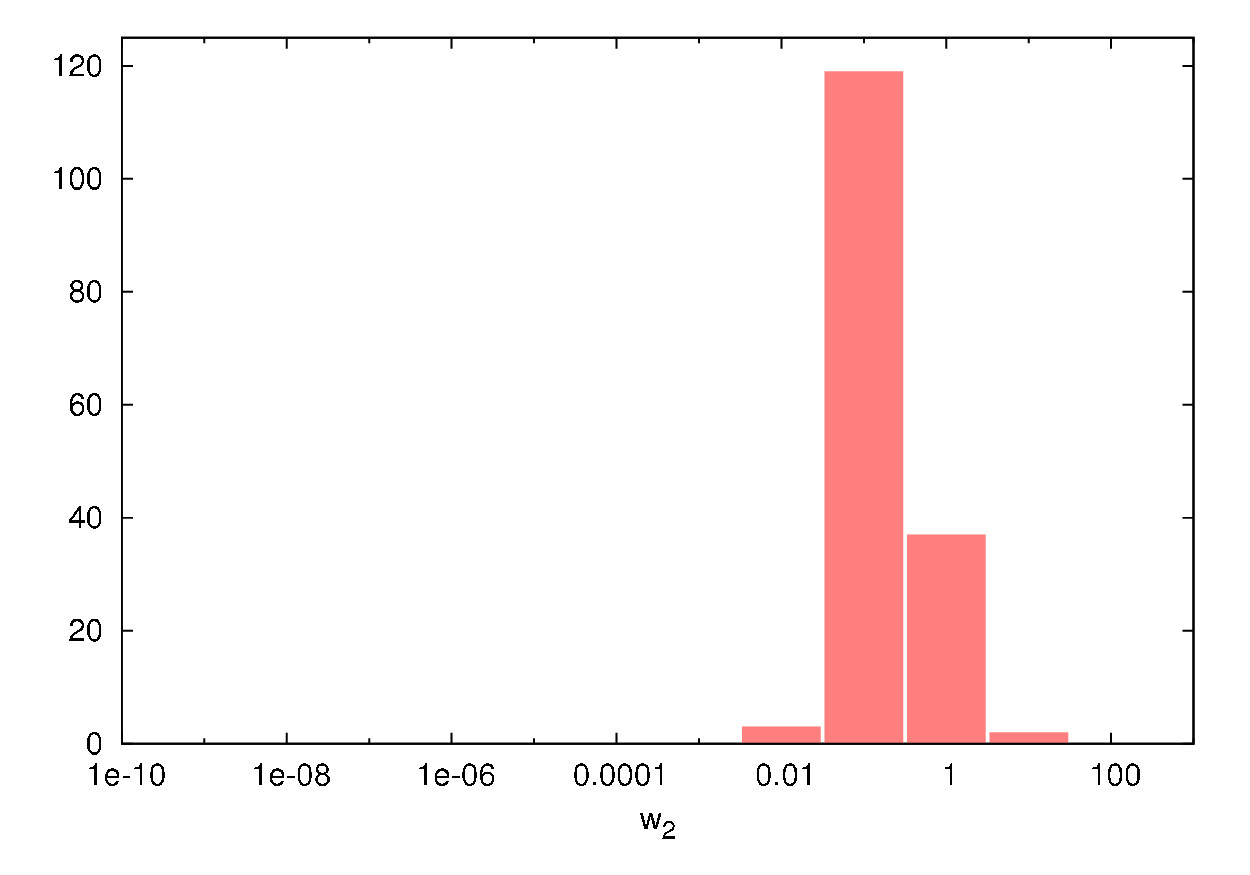}
\includegraphics[width=0.49\columnwidth]{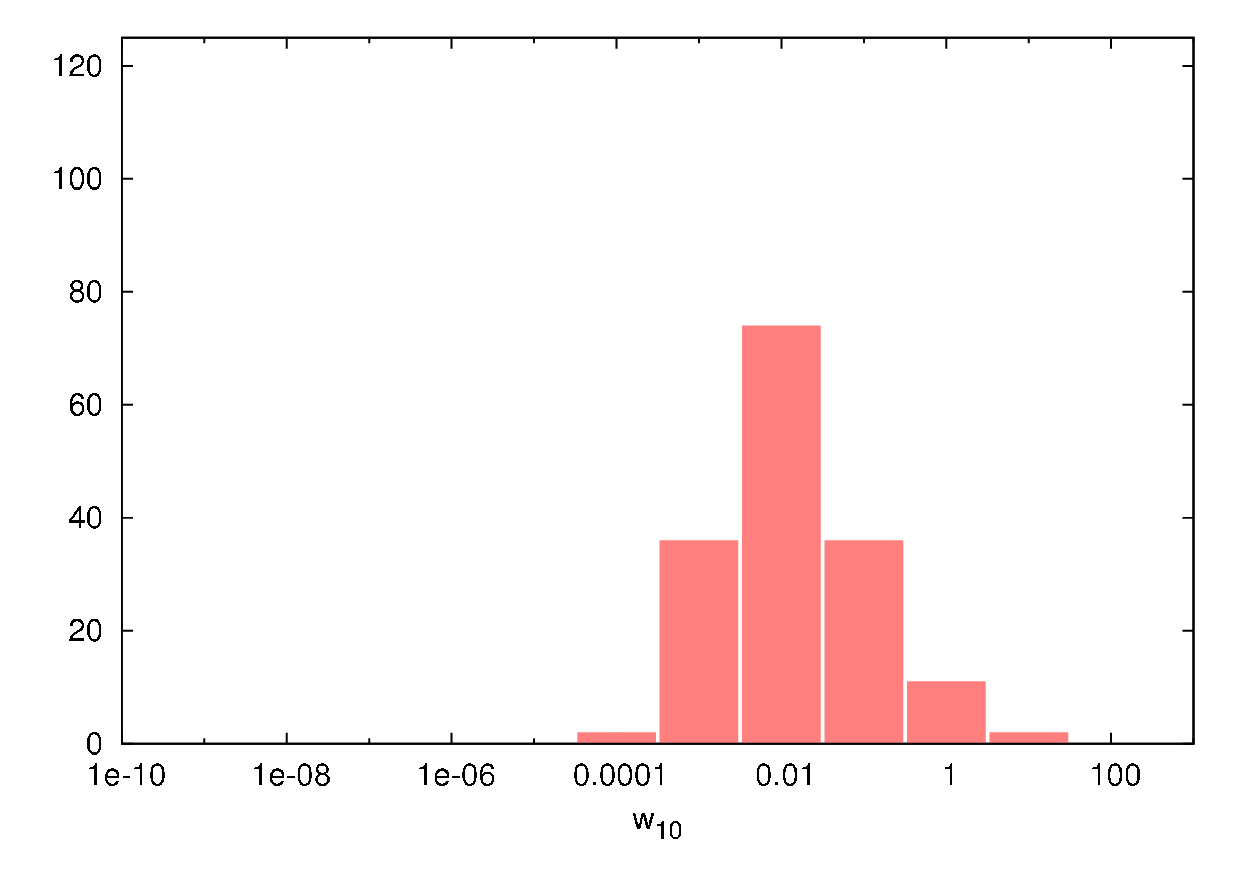}\\
\includegraphics[width=0.49\columnwidth]{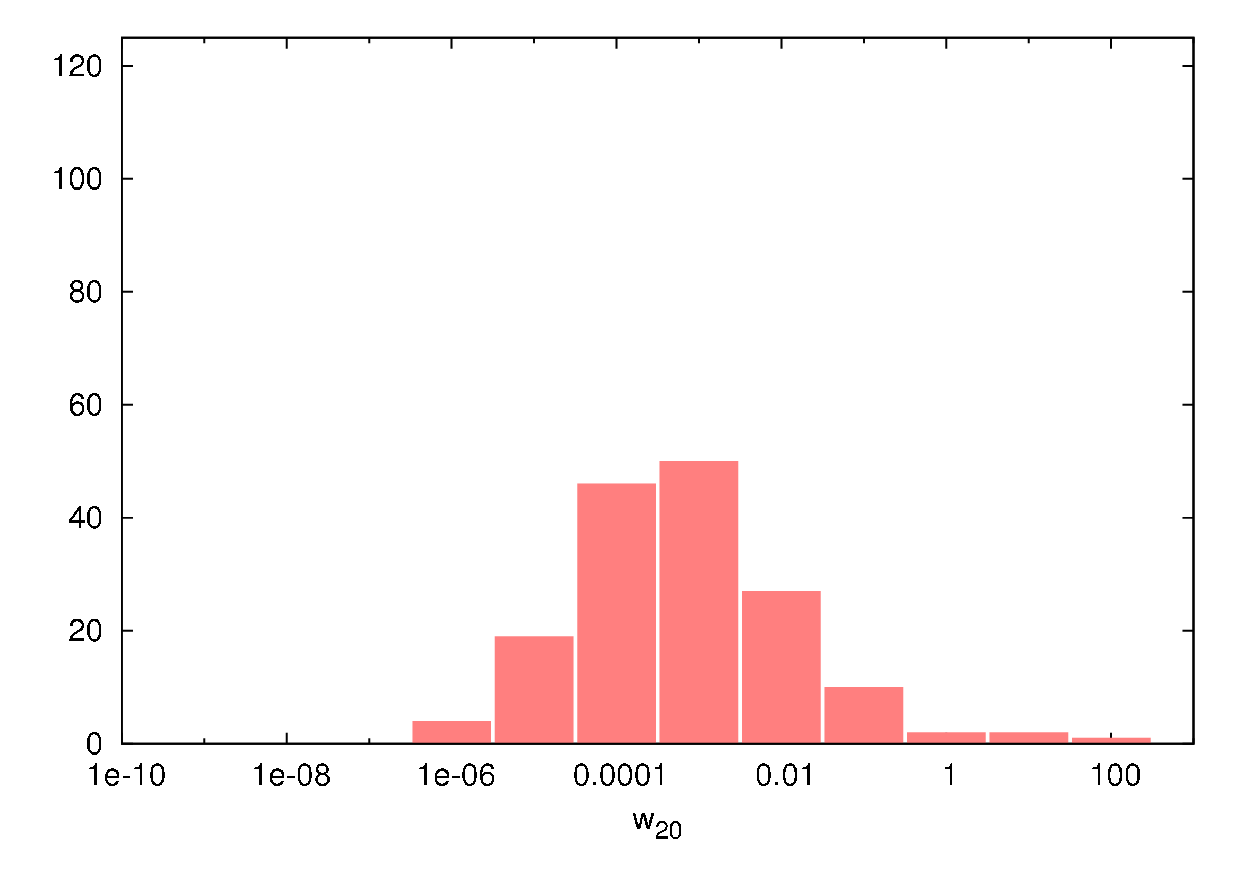}
\includegraphics[width=0.49\columnwidth]{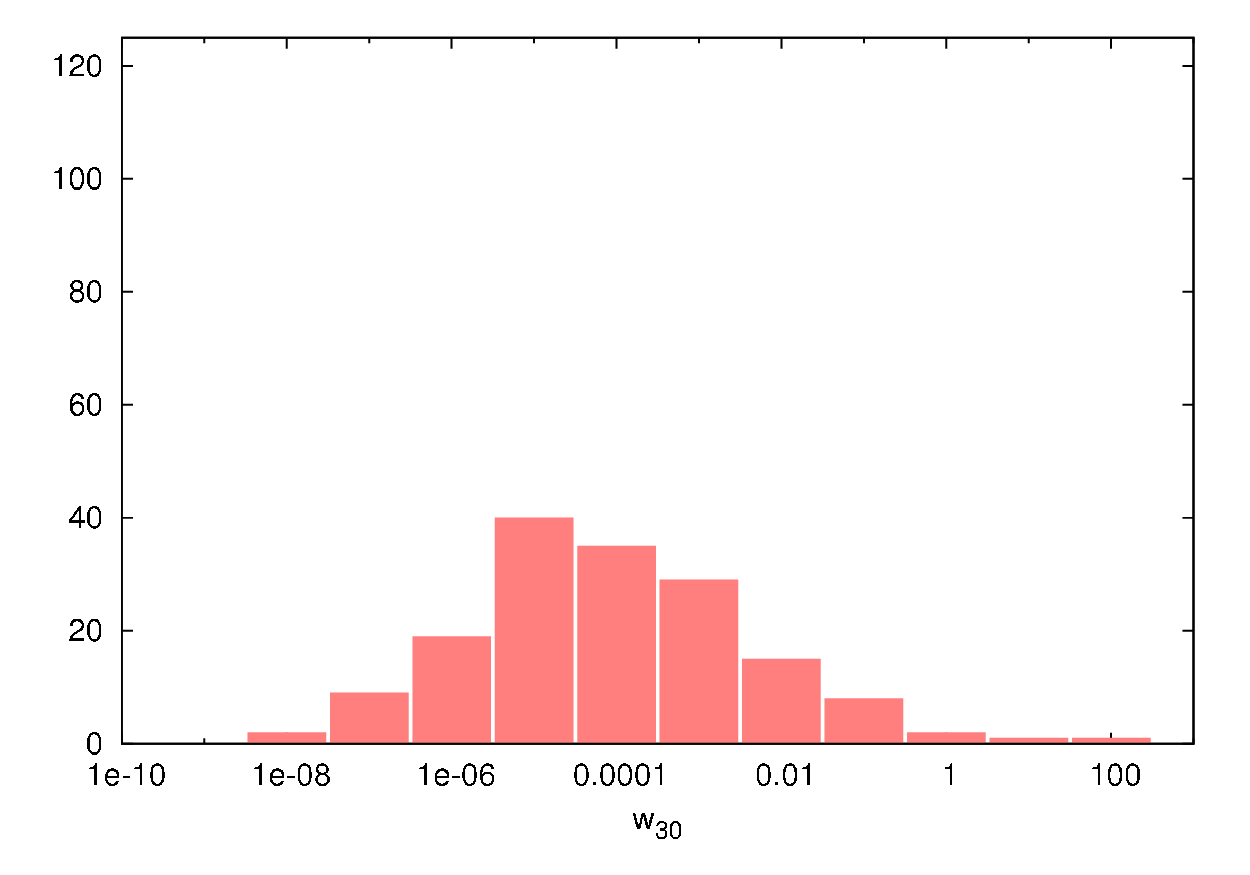}
\caption{Histograms of the values $\overline{w}_k$
for $k=2$ (top left), $k=10$ (top right), $k=20$ (bottom left) 
and $k=30$ (bottom right) from a set of 161 gauge field configurations.}\label{wxx}
\end{figure}

We can bring \eq{OUwn} in a form more similar to the unweighted case
by defining the ratio
\be
	\overline{w}_k\left[U_n\right]\equiv \frac{w_k\left[U_n\right]}{\sum_{n}w_k\left[U_n\right]}\cdot N\;.
\ee
We can rewrite \eq{OUwn} to obtain
\be
	\left<\bigO\left[U\right]\right>_{w_k} \approx \frac{1}{N}\sum_n \bigO\left[U_n\right] \overline{w}_k\left[U_n\right]\,,
\ee
which equals the unweighted case up to the
the factors $\overline{w}_k\left[U_n\right]$, which multiply the observable $\bigO\left[U_n\right]$
on each configuration that enters into the Monte Carlo integration.

In \fig{wxx} we show the distribution of the values $\overline{w}_k$
for $k=2, 10, 20, 30$ from our set of 161 gauge field configurations.
Unfortunately, this shows that for truncation levels $k\geq 20$, which are the level
of interest for the restoration of the chiral symmetry (see \fig{fig:fitted_masses_mesons}),
the Monte Carlo sum is highly dominated by very few gauge configurations.
Therefore, we would need many more gauge field configurations in order
to obtain the same statistical quality as before.
There is a chance, however, that the distribution of the weight factors is narrower
for overlap fermions due to the strict circular distribution of the eigenvalues of 
the overlap operator in the complex plane, opposed to the eigenvalues of the CI  that are 
spread with respect to the circle.

\section{Conclusions}\label{sec:conclusions}
We find that our procedure of leaving out the lowest-lying part of the 
Dirac eigenspectrum
does not crucially violate the locality of the theory. We have shown in \Sec{sec:locality}
that the nonlocal contributions to the truncated Dirac operator are of the order of one million
times smaller than the local contributions, and moreover, fall exponentially with the distance
at finite lattice spacing. 

The effects of Dirac low-mode removal on the quarks are twofold: 
first, the dynamical mass generation of the quarks
vanishes, signaling the restoration of the chiral symmetry.
Secondly, the average momentum of the quarks is increased compared
with the full theory which accounts for the larger hadron masses
and is responsible for a partial effective chiral restoration at low truncation levels.

Lastly, we introduced a low-mode reweighting procedure to extend our approximation to the sea quark
sector. The broad shape of the distribution of the weight factors renders the scheme impractical
for CI fermions. 

\begin{acknowledgments}
We thank Pedro Bicudo for helpful discussions.
M.S. is supported by the Research Executive 
Agency (REA) of the European Union under Grant Agreement 
PITN-GA-2009-238353 (ITN STRONGnet).

\end{acknowledgments}


\providecommand{\href}[2]{#2}\begingroup\raggedright\endgroup

\end{document}